# Measurement of the Branching Fraction for $D_s^+ \to \tau^+\nu_\tau$ and Extraction of the Decay Constant $f_{D_s}$

J. P. Lees,[1] V. Poireau,[1] E. Prencipe,[1] V. Tisserand,[1] J. Garra Tico,[2] E. Grauges,[2] M. Martinelli[ab],[3] A. Palano[ab],[3] M. Pappagallo[ab],[3] G. Eigen,[4] B. Stugu,[4] L. Sun,[4] M. Battaglia,[5] D. N. Brown,[5] B. Hooberman,[5] L. T. Kerth,[5] Yu. G. Kolomensky,[5] G. Lynch,[5] I. L. Osipenkov,[5] T. Tanabe,[5] C. M. Hawkes,[6] N. Soni,[6] A. T. Watson,[6] H. Koch,[7] T. Schroeder,[7] D. J. Asgeirsson,[8] C. Hearty,[8] T. S. Mattison,[8] J. A. McKenna,[8] M. Barrett,[9] A. Khan,[9] A. Randle-Conde,[9] V. E. Blinov,[10] A. R. Buzykaev,[10] V. P. Druzhinin,[10] V. B. Golubev,[10] A. P. Onuchin,[10] S. I. Serednyakov,[10] Yu. I. Skovpen,[10] E. P. Solodov,[10] K. Yu. Todyshev,[10] A. N. Yushkov,[10] M. Bondioli,[11] S. Curry,[11] D. Kirkby,[11] A. J. Lankford,[11] P. Lund,[11] M. Mandelkern,[11] E. C. Martin,[11] D. P. Stoker,[11] H. Atmacan,[12] J. W. Gary,[12] F. Liu,[12] O. Long,[12] G. M. Vitug,[12] Z. Yasin,[12] V. Sharma,[13] C. Campagnari,[14] T. M. Hong,[14] D. Kovalskyi,[14] J. D. Richman,[14] A. M. Eisner,[15] C. A. Heusch,[15] J. Kroseberg,[15] W. S. Lockman,[15] A. J. Martinez,[15] T. Schalk,[15] B. A. Schumm,[15] A. Seiden,[15] L. O. Winstrom,[15] C. H. Cheng,[16] D. A. Doll,[16] B. Echenard,[16] D. G. Hitlin,[16] P. Ongmongkolkul,[16] F. C. Porter,[16] A. Y. Rakitin,[16] R. Andreassen,[17] M. S. Dubrovin,[17] G. Mancinelli,[17] B. T. Meadows,[17] M. D. Sokoloff,[17] P. C. Bloom,[18] W. T. Ford,[18] A. Gaz,[18] J. F. Hirschauer,[18] M. Nagel,[18] U. Nauenberg,[18] J. G. Smith,[18] S. R. Wagner,[18] R. Ayad,[19,*] W. H. Toki,[19] E. Feltresi,[20] A. Hauke,[20] H. Jasper,[20] T. M. Karbach,[20] J. Merkel,[20] A. Petzold,[20] B. Spaan,[20] K. Wacker,[20] M. J. Kobel,[21] K. R. Schubert,[21] R. Schwierz,[21] D. Bernard,[22] M. Verderi,[22] P. J. Clark,[23] S. Playfer,[23] J. E. Watson,[23] M. Andreotti[ab],[24] D. Bettoni[a],[24] C. Bozzi[a],[24] R. Calabrese[ab],[24] A. Cecchi[ab],[24] G. Cibinetto[ab],[24] E. Fioravanti[ab],[24] P. Franchini[ab],[24] E. Luppi[ab],[24] M. Munerato[ab],[24] M. Negrini[ab],[24] A. Petrella[ab],[24] L. Piemontese[a],[24] V. Santoro[ab],[24] R. Baldini-Ferroli,[25] A. Calcaterra,[25] R. de Sangro,[25] G. Finocchiaro,[25] M. Nicolaci,[25] S. Pacetti,[25] P. Patteri,[25] I. M. Peruzzi,[25,†] M. Piccolo,[25] M. Rama,[25] A. Zallo,[25] R. Contri[ab],[26] E. Guido[ab],[26] M. Lo Vetere[ab],[26] M. R. Monge[ab],[26] S. Passaggio[a],[26] C. Patrignani[ab],[26] E. Robutti[a],[26] S. Tosi[ab],[26] B. Bhuyan,[27] M. Morii,[28] A. Adametz,[29] J. Marks,[29] S. Schenk,[29] U. Uwer,[29] F. U. Bernlochner,[30] H. M. Lacker,[30] T. Lueck,[30] A. Volk,[30] P. D. Dauncey,[31] M. Tibbetts,[31] P. K. Behera,[32] U. Mallik,[32] C. Chen,[33] J. Cochran,[33] H. B. Crawley,[33] L. Dong,[33] W. T. Meyer,[33] S. Prell,[33] E. I. Rosenberg,[33] A. E. Rubin,[33] Y. Y. Gao,[34] A. V. Gritsan,[34] Z. J. Guo,[34] N. Arnaud,[35] M. Davier,[35] D. Derkach,[35] J. Firmino da Costa,[35] G. Grosdidier,[35] F. Le Diberder,[35] A. M. Lutz,[35] B. Malaescu,[35] P. Roudeau,[35] M. H. Schune,[35] J. Serrano,[35] V. Sordini,[35,‡] A. Stocchi,[35] L. Wang,[35] G. Wormser,[35] D. J. Lange,[36] D. M. Wright,[36] I. Bingham,[37] J. P. Burke,[37] C. A. Chavez,[37] J. R. Fry,[37] E. Gabathuler,[37] R. Gamet,[37] D. E. Hutchcroft,[37] D. J. Payne,[37] C. Touramanis,[37] A. J. Bevan,[38] F. Di Lodovico,[38] R. Sacco,[38] M. Sigamani,[38] G. Cowan,[39] S. Paramesvaran,[39] A. C. Wren,[39] D. N. Brown,[40] C. L. Davis,[40] A. G. Denig,[41] M. Fritsch,[41] W. Gradl,[41] A. Hafner,[41] K. E. Alwyn,[42] D. Bailey,[42] R. J. Barlow,[42] G. Jackson,[42] G. D. Lafferty,[42] T. J. West,[42] J. Anderson,[43] A. Jawahery,[43] D. A. Roberts,[43] G. Simi,[43] J. M. Tuggle,[43] C. Dallapiccola,[44] E. Salvati,[44] R. Cowan,[45] D. Dujmic,[45] P. H. Fisher,[45] G. Sciolla,[45] R. K. Yamamoto,[45] M. Zhao,[45] P. M. Patel,[46] S. H. Robertson,[46] M. Schram,[46] P. Biassoni[ab],[47] A. Lazzaro[ab],[47] V. Lombardo[a],[47] F. Palombo[ab],[47] S. Stracka[ab],[47] L. Cremaldi,[48] R. Godang,[48,§] R. Kroeger,[48] P. Sonnek,[48] D. J. Summers,[48] H. W. Zhao,[48] X. Nguyen,[49] M. Simard,[49] P. Taras,[49] G. De Nardo[ab],[50] D. Monorchio[ab],[50] G. Onorato[ab],[50] C. Sciacca[ab],[50] G. Raven,[51] H. L. Snoek,[51] C. P. Jessop,[52] K. J. Knoepfel,[52] J. M. LoSecco,[52] W. F. Wang,[52] L. A. Corwin,[53] K. Honscheid,[53] R. Kass,[53] J. P. Morris,[53] A. M. Rahimi,[53] S. J. Sekula,[53] N. L. Blount,[54] J. Brau,[54] R. Frey,[54] O. Igonkina,[54] J. A. Kolb,[54] M. Lu,[54] R. Rahmat,[54] N. B. Sinev,[54] D. Strom,[54] J. Strube,[54] E. Torrence,[54] G. Castelli[ab],[55] N. Gagliardi[ab],[55] M. Margoni[ab],[55] M. Morandin[a],[55] M. Posocco[a],[55] M. Rotondo[a],[55] F. Simonetto[ab],[55] R. Stroili[ab],[55] P. del Amo Sanchez,[56] E. Ben-Haim,[56] G. R. Bonneaud,[56] H. Briand,[56] J. Chauveau,[56] O. Hamon,[56] Ph. Leruste,[56] G. Marchiori,[56] J. Ocariz,[56] A. Perez,[56] J. Prendki,[56] S. Sitt,[56] M. Biasini[ab],[57] E. Manoni[ab],[57] C. Angelini[ab],[58] G. Batignani[ab],[58] S. Bettarini[ab],[58] G. Calderini[ab],[58,¶] M. Carpinelli[ab],[58,**] A. Cervelli[ab],[58] F. Forti[ab],[58] M. A. Giorgi[ab],[58] A. Lusiani[ac],[58] N. Neri[ab],[58] E. Paoloni[ab],[58] G. Rizzo[ab],[58] J. J. Walsh[a],[58] D. Lopes Pegna,[59] C. Lu,[59] J. Olsen,[59] A. J. S. Smith,[59] A. V. Telnov,[59] F. Anulli[a],[60] E. Baracchini[ab],[60] G. Cavoto[a],[60] R. Faccini[ab],[60] F. Ferrarotto[a],[60] F. Ferroni[ab],[60] M. Gaspero[ab],[60] P. D. Jackson[a],[60] L. Li Gioi[a],[60] M. A. Mazzoni[a],[60] G. Piredda[a],[60] F. Renga[ab],[60]

M. Ebert,[61] T. Hartmann,[61] T. Leddig,[61] H. Schröder,[61] R. Waldi,[61] T. Adye,[62] B. Franek,[62] E. O. Olaiya,[62] F. F. Wilson,[62] S. Emery,[63] G. Hamel de Monchenault,[63] G. Vasseur,[63] Ch. Yèche,[63] M. Zito,[63] M. T. Allen,[64] D. Aston,[64] D. J. Bard,[64] R. Bartoldus,[64] J. F. Benitez,[64] C. Cartaro,[64] R. Cenci,[64] J. P. Coleman,[64] M. R. Convery,[64] J. C. Dingfelder,[64] J. Dorfan,[64] G. P. Dubois-Felsmann,[64] W. Dunwoodie,[64] R. C. Field,[64] M. Franco Sevilla,[64] B. G. Fulsom,[64] A. M. Gabareen,[64] M. T. Graham,[64] P. Grenier,[64] C. Hast,[64] W. R. Innes,[64] J. Kaminski,[64] M. H. Kelsey,[64] H. Kim,[64] P. Kim,[64] M. L. Kocian,[64] D. W. G. S. Leith,[64] S. Li,[64] B. Lindquist,[64] S. Luitz,[64] V. Luth,[64] H. L. Lynch,[64] D. B. MacFarlane,[64] H. Marsiske,[64] R. Messner,[64, ††] D. R. Muller,[64] H. Neal,[64] S. Nelson,[64] C. P. O'Grady,[64] I. Ofte,[64] M. Perl,[64] B. N. Ratcliff,[64] A. Roodman,[64] A. A. Salnikov,[64] R. H. Schindler,[64] J. Schwiening,[64] A. Snyder,[64] D. Su,[64] M. K. Sullivan,[64] K. Suzuki,[64] S. K. Swain,[64] J. M. Thompson,[64] J. Va'vra,[64] A. P. Wagner,[64] M. Weaver,[64] C. A. West,[64] W. J. Wisniewski,[64] M. Wittgen,[64] D. H. Wright,[64] H. W. Wulsin,[64] A. K. Yarritu,[64] C. C. Young,[64] V. Ziegler,[64] X. R. Chen,[65] H. Liu,[65] W. Park,[65] M. V. Purohit,[65] R. M. White,[65] J. R. Wilson,[65] M. Bellis,[66] P. R. Burchat,[66] A. J. Edwards,[66] T. S. Miyashita,[66] S. Ahmed,[67] M. S. Alam,[67] J. A. Ernst,[67] B. Pan,[67] M. A. Saeed,[67] S. B. Zain,[67] N. Guttman,[68] A. Soffer,[68] S. M. Spanier,[69] B. J. Wogsland,[69] R. Eckmann,[70] J. L. Ritchie,[70] A. M. Ruland,[70] C. J. Schilling,[70] R. F. Schwitters,[70] B. C. Wray,[70] J. M. Izen,[71] X. C. Lou,[71] F. Bianchi[ab],[72] D. Gamba[ab],[72] M. Pelliccioni[ab],[72] M. Bomben[ab],[73] G. Della Ricca[ab],[73] L. Lanceri[ab],[73] L. Vitale[ab],[73] V. Azzolini,[74] N. Lopez-March,[74] F. Martinez-Vidal,[74] D. A. Milanes,[74] A. Oyanguren,[74] J. Albert,[75] Sw. Banerjee,[75] H. H. F. Choi,[75] K. Hamano,[75] G. J. King,[75] R. Kowalewski,[75] M. J. Lewczuk,[75] I. M. Nugent,[75] J. M. Roney,[75] R. J. Sobie,[75] T. J. Gershon,[76] P. F. Harrison,[76] J. Ilic,[76] T. E. Latham,[76] G. B. Mohanty,[76] E. M. T. Puccio,[76] H. R. Band,[77] X. Chen,[77] S. Dasu,[77] K. T. Flood,[77] Y. Pan,[77] R. Prepost,[77] C. O. Vuosalo,[77] and S. L. Wu[77]

(The *BABAR* Collaboration)

[1]*Laboratoire d'Annecy-le-Vieux de Physique des Particules (LAPP), Université de Savoie, CNRS/IN2P3, F-74941 Annecy-Le-Vieux, France*

[2]*Universitat de Barcelona, Facultat de Fisica, Departament ECM, E-08028 Barcelona, Spain*

[3]*INFN Sezione di Bari[a]; Dipartimento di Fisica, Università di Bari[b], I-70126 Bari, Italy*

[4]*University of Bergen, Institute of Physics, N-5007 Bergen, Norway*

[5]*Lawrence Berkeley National Laboratory and University of California, Berkeley, California 94720, USA*

[6]*University of Birmingham, Birmingham, B15 2TT, United Kingdom*

[7]*Ruhr Universität Bochum, Institut für Experimentalphysik 1, D-44780 Bochum, Germany*

[8]*University of British Columbia, Vancouver, British Columbia, Canada V6T 1Z1*

[9]*Brunel University, Uxbridge, Middlesex UB8 3PH, United Kingdom*

[10]*Budker Institute of Nuclear Physics, Novosibirsk 630090, Russia*

[11]*University of California at Irvine, Irvine, California 92697, USA*

[12]*University of California at Riverside, Riverside, California 92521, USA*

[13]*University of California at San Diego, La Jolla, California 92093, USA*

[14]*University of California at Santa Barbara, Santa Barbara, California 93106, USA*

[15]*University of California at Santa Cruz, Institute for Particle Physics, Santa Cruz, California 95064, USA*

[16]*California Institute of Technology, Pasadena, California 91125, USA*

[17]*University of Cincinnati, Cincinnati, Ohio 45221, USA*

[18]*University of Colorado, Boulder, Colorado 80309, USA*

[19]*Colorado State University, Fort Collins, Colorado 80523, USA*

[20]*Technische Universität Dortmund, Fakultät Physik, D-44221 Dortmund, Germany*

[21]*Technische Universität Dresden, Institut für Kern- und Teilchenphysik, D-01062 Dresden, Germany*

[22]*Laboratoire Leprince-Ringuet, CNRS/IN2P3, Ecole Polytechnique, F-91128 Palaiseau, France*

[23]*University of Edinburgh, Edinburgh EH9 3JZ, United Kingdom*

[24]*INFN Sezione di Ferrara[a]; Dipartimento di Fisica, Università di Ferrara[b], I-44100 Ferrara, Italy*

[25]*INFN Laboratori Nazionali di Frascati, I-00044 Frascati, Italy*

[26]*INFN Sezione di Genova[a]; Dipartimento di Fisica, Università di Genova[b], I-16146 Genova, Italy*

[27]*Indian Institute of Technology Guwahati, Guwahati, Assam, 781 039, India*

[28]*Harvard University, Cambridge, Massachusetts 02138, USA*

[29]*Universität Heidelberg, Physikalisches Institut, Philosophenweg 12, D-69120 Heidelberg, Germany*

[30]*Humboldt-Universität zu Berlin, Institut für Physik, Newtonstr. 15, D-12489 Berlin, Germany*

[31]*Imperial College London, London, SW7 2AZ, United Kingdom*

[32]*University of Iowa, Iowa City, Iowa 52242, USA*

[33]*Iowa State University, Ames, Iowa 50011-3160, USA*

[34]*Johns Hopkins University, Baltimore, Maryland 21218, USA*

[35]*Laboratoire de l'Accélérateur Linéaire, IN2P3/CNRS et Université Paris-Sud 11, Centre Scientifique d'Orsay, B. P. 34, F-91898 Orsay Cedex, France*

[36]Lawrence Livermore National Laboratory, Livermore, California 94550, USA
[37]University of Liverpool, Liverpool L69 7ZE, United Kingdom
[38]Queen Mary, University of London, London, E1 4NS, United Kingdom
[39]University of London, Royal Holloway and Bedford New College, Egham, Surrey TW20 0EX, United Kingdom
[40]University of Louisville, Louisville, Kentucky 40292, USA
[41]Johannes Gutenberg-Universität Mainz, Institut für Kernphysik, D-55099 Mainz, Germany
[42]University of Manchester, Manchester M13 9PL, United Kingdom
[43]University of Maryland, College Park, Maryland 20742, USA
[44]University of Massachusetts, Amherst, Massachusetts 01003, USA
[45]Massachusetts Institute of Technology, Laboratory for Nuclear Science, Cambridge, Massachusetts 02139, USA
[46]McGill University, Montréal, Québec, Canada H3A 2T8
[47]INFN Sezione di Milano[a]; Dipartimento di Fisica, Università di Milano[b], I-20133 Milano, Italy
[48]University of Mississippi, University, Mississippi 38677, USA
[49]Université de Montréal, Physique des Particules, Montréal, Québec, Canada H3C 3J7
[50]INFN Sezione di Napoli[a]; Dipartimento di Scienze Fisiche,
Università di Napoli Federico II[b], I-80126 Napoli, Italy
[51]NIKHEF, National Institute for Nuclear Physics and High Energy Physics, NL-1009 DB Amsterdam, The Netherlands
[52]University of Notre Dame, Notre Dame, Indiana 46556, USA
[53]Ohio State University, Columbus, Ohio 43210, USA
[54]University of Oregon, Eugene, Oregon 97403, USA
[55]INFN Sezione di Padova[a]; Dipartimento di Fisica, Università di Padova[b], I-35131 Padova, Italy
[56]Laboratoire de Physique Nucléaire et de Hautes Energies,
IN2P3/CNRS, Université Pierre et Marie Curie-Paris6,
Université Denis Diderot-Paris7, F-75252 Paris, France
[57]INFN Sezione di Perugia[a]; Dipartimento di Fisica, Università di Perugia[b], I-06100 Perugia, Italy
[58]INFN Sezione di Pisa[a]; Dipartimento di Fisica,
Università di Pisa[b]; Scuola Normale Superiore di Pisa[c], I-56127 Pisa, Italy
[59]Princeton University, Princeton, New Jersey 08544, USA
[60]INFN Sezione di Roma[a]; Dipartimento di Fisica,
Università di Roma La Sapienza[b], I-00185 Roma, Italy
[61]Universität Rostock, D-18051 Rostock, Germany
[62]Rutherford Appleton Laboratory, Chilton, Didcot, Oxon, OX11 0QX, United Kingdom
[63]CEA, Irfu, SPP, Centre de Saclay, F-91191 Gif-sur-Yvette, France
[64]SLAC National Accelerator Laboratory, Stanford, California 94309 USA
[65]University of South Carolina, Columbia, South Carolina 29208, USA
[66]Stanford University, Stanford, California 94305-4060, USA
[67]State University of New York, Albany, New York 12222, USA
[68]Tel Aviv University, School of Physics and Astronomy, Tel Aviv, 69978, Israel
[69]University of Tennessee, Knoxville, Tennessee 37996, USA
[70]University of Texas at Austin, Austin, Texas 78712, USA
[71]University of Texas at Dallas, Richardson, Texas 75083, USA
[72]INFN Sezione di Torino[a]; Dipartimento di Fisica Sperimentale, Università di Torino[b], I-10125 Torino, Italy
[73]INFN Sezione di Trieste[a]; Dipartimento di Fisica, Università di Trieste[b], I-34127 Trieste, Italy
[74]IFIC, Universitat de Valencia-CSIC, E-46071 Valencia, Spain
[75]University of Victoria, Victoria, British Columbia, Canada V8W 3P6
[76]Department of Physics, University of Warwick, Coventry CV4 7AL, United Kingdom
[77]University of Wisconsin, Madison, Wisconsin 53706, USA

The branching fraction for the decay $D_s^+ \to \tau^+\nu_\tau$, with $\tau^+ \to e^+\nu_e\overline{\nu}_\tau$, is measured using a data sample corresponding to an integrated luminosity of 427 $\mathrm{fb}^{-1}$ collected at center of mass energies near 10.58 GeV with the *BABAR* detector at the PEP-II asymmetric-energy $e^+e^-$ collider at SLAC. In the process $e^+e^- \to c\overline{c} \to D_s^{*+}\overline{D}_{\mathrm{TAG}}\overline{K}X$, the $D_s^{*+}$ meson is reconstructed as a missing particle, and the subsequent decay $D_s^{*+} \to D_s^+\gamma$ yields an inclusive $D_s^+$ data sample. Here $\overline{D}_{\mathrm{TAG}}$ refers to a fully reconstructed hadronic $\overline{D}$ decay, $\overline{K}$ is a $K^-$ or $\overline{K}^0$, and $X$ stands for any number of charged or neutral pions. The decay $D_s^+ \to K_S^0K^+$ is isolated also, and from ratio of event yields and known branching fractions, $\mathcal{B}(D_s^+ \to \tau^+\nu_\tau) = (4.5\pm0.5\pm0.4\pm0.3)\%$ is determined. The pseudoscalar decay constant is extracted to be $f_{D_s} = (233\pm13\pm10\pm7)$ MeV, where the first uncertainty is statistical, the second is systematic, and the third results from the uncertainties on the external measurements used as input to the calculation.

The $D_s^+$ meson can decay purely leptonically via annihilation of the $c$ and $\overline{s}$ quarks to a virtual $W^+$ boson which decays to a lepton pair. These decays provide a clean probe of the pseudoscalar meson decay constant $f_{D_s}$, which describes the amplitude for the $c$ and $\overline{s}$ quarks to have zero spatial separation within the meson, a nec-essary condition for the annihilation to take place. In the Standard Model (SM), ignoring radiative processes, the total width is

$$\Gamma(D_s^+ \to \ell^+\nu_\ell) = \frac{G_F^2}{8\pi} M_{D_s^+}^3 \left(\frac{m_\ell}{M_{D_s^+}}\right)^2 \left(1 - \frac{m_\ell^2}{M_{D_s^+}^2}\right)^2 |V_{cs}|^2 f_{D_s}^2, \quad (1)$$

where $M_{D_s^+}$ and $m_\ell$ are the $D_s^+$ and lepton masses, respectively, $G_F$ is the Fermi coupling constant, $|V_{cs}|$ is the magnitude of the Cabibbo-Kobayashi-Maskawa (CKM) matrix element that characterizes the coupling of the weak charged current to the $c$ and $\overline{s}$ quarks [1].

The leptonic decay of the $D_s^+$ meson is helicity-suppressed because it has zero spin, so that the final state neutrino and lepton must combine to form a spin-0 state. Consequently, the left-handed neutrino forces the anti-lepton to be left-handed, thus suppressing the decay rate by the factor $m_\ell^2/M_{D_s^+}^2$. The net effect of helicity and phase space factors results in large differences in the leptonic branching fractions of the $D_s^+$ meson. The branching fractions for $D_s^+$ decays to $\bar{l}\nu_\ell$, where $\bar{l} = e^+$, $\mu^+$, $\tau^+$, are roughly $2 \times 10^{-5}$: $1 : 10$ in proportion. The large branching fraction for the $\tau^+$ decay mode motivates the use of the decay sequence $D_s^+ \to \tau^+\nu_\tau$, $\tau^+ \to e^+\nu_e\overline{\nu}_\tau$ in this analysis. The signal branching fraction $\mathcal{B}(D_s^+ \to \tau^+\nu_\tau)$ relative to the well measured branching fraction $\mathcal{B}(D_s^+ \to K_S^0 K^+) = (1.49\pm0.09)\%$ [2] is determined and used to extract the decay constant $f_{D_s}$.

In the context of the SM, predictions for meson decay constants can be obtained from QCD lattice calculations [3–8]. The most precise theoretical prediction for $f_{D_s}$, which uses unquenched lattice QCD, is (241±3) MeV [5]. The most precise measurement of the branching fraction for $D_s^+ \to \tau^+\nu_\tau$ ($\tau^+ \to e^+\nu_e\overline{\nu}_\tau$) yields $\mathcal{B}(D_s^+ \to \tau^+\nu_\tau)$= (5.30±0.47±0.22)% [10] and the value $f_{D_s}$ = (252.5±11.1±5.2) MeV. Decay constants of $D$ and $B$ mesons enter into calculations of hadronic matrix elements for several key processes and their theoretical predictions. For instance the calculation of $B\overline{B}$ mixing requires knowledge of $f_B$. While leptonic decay of the $B$ meson is heavily CKM suppressed, leptonic decay of $D_s^+$ meson is CKM favored and thus resulting more precise measurements of $f_{D_s}$ can be used to validate the lattice QCD calculations that are applicable to $B$-meson decay. Several models involving physics beyond the SM can induce a difference between the theoretical prediction and the measured value. These include a two-Higgs doublet model [12] and a model incorporating two leptoquarks [13]. It is important to have high precision determinations of $f_{D_s}$, both from experiment and theory, in order to discover or constrain effects of physics beyond the SM. The Particle Data Group gives a world average of $f_{D_s}$ = (273±10) MeV [9] but this does not include the most recent results [10, 11].

Measuring the branching fraction for $D_s^+ \to \tau^+\nu_\tau$ requires knowledge of the total number of $D_s^+$ mesons in the parent analysis sample. Alternatively, the branching fraction can be measured relative to that for a $D_s^+$ decay mode with well-known branching fraction, with the latter then used to obtain $\mathcal{B}(D_s^+ \to \tau^+\nu_\tau)$; this is the procedure followed in the present analysis. The decay mode $D_s^+ \to \phi\pi^+$ has been used often as a normalization mode. However this is somewhat problematic, since determination of the branching fraction for this decay requires a Dalitz plot analysis of the $D_s^+ \to K^+K^-\pi^+$ process. A description of the Dalitz plot intensity distribution incorporates contributions from other quasi-two-body amplitudes such as $\overline{K}^*(892)^0K^+$, $\overline{K}^*(1430)^0K^+$ and $f_0(980)\pi^+$. The contributions of these other decay modes to the specific mass range used to define the $\phi \to K^+K^-$ rate have to be taken into account. These depend on the mass and width values of the resonances and their interference effects, as well as the mass resolution of the experiment. [2, 14]. For these reasons, the decay mode $D_s^+ \to K_S^0 K^+$ is chosen instead as reference mode in the present analysis. Its branching fraction is quite well known, and the branching fraction for $D_s^+ \to \tau^+\nu_\tau$ can then be expressed as:

$$\frac{\mathcal{B}(D_s^+ \to \tau^+\nu_\tau)}{\mathcal{B}(D_s^+ \to K_S^0 K^+)} = \frac{\mathcal{B}(K_S^0 \to \pi^+\pi^-)}{\mathcal{B}(\tau^+ \to e^+\nu_e\overline{\nu}_\tau)} \frac{(N_S)^{\tau\nu_\tau}}{(N_S)^{K_S^0 K^+}} \frac{\epsilon^{K_S^0 K^+}}{\epsilon^{\tau\nu_\tau}}, \quad (2)$$

where $N_S$ and $\epsilon$ refer to the number of events and total efficiency for the signal and the normalizing decay modes. The values of the branching fractions used for $K_S^0 \to \pi^+\pi^-$ and $\tau^+ \to e^+\nu_e\overline{\nu}_\tau$ are (69.20±0.05)% and (17.85±0.05)%, respectively [9].

This analysis uses an integrated luminosity of 427 fb$^{-1}$ for $e^+e^-$ collisions at center of mass (CM) energies near 10.58 GeV, corresponding to the production of approximately 554 million $c\overline{c}$ events. The data were collected with the *BABAR* detector at the SLAC PEP-II asymmetric-energy collider. The *BABAR* detector is described in detail in Refs. [15, 16]. Charged-particle momenta are measured with a 5-layer, double-sided silicon vertex tracker (SVT) and a 40-layer drift chamber (DCH) embedded in the 1.5-T magnetic field. A calorimeter consisting of 6580 CsI(Tl) crystals is used to measure electromagnetic energy. Charged pions and kaons are identified by a ring imaging Cherenkov detector (DIRC) and by their specific ionization loss in the SVT and DCH. Muons penetrating the solenoid are detected in the instrumented magnet flux return.

The $D_s^+ \to \tau^+\nu_\tau$ branching fraction measurement is carried out via the $D_s^{*+}$ production process $e^+e^- \to c\overline{c} \to D_s^{*+}\overline{D}_{\rm TAG}\overline{K}^{0,-}X$, and the subsequent decay $D_s^{*+} \to D_s^+\gamma$. Here, $\overline{D}_{\rm TAG}$ is a fully reconstructed hadronic $\overline{D}$ meson decay, required to suppress the large background from non-charm continuum $q\overline{q}$ pair production; $X$ represents a set of any number of pions ($\pi^0$ and $\pi^\pm$) produced in the $c\overline{c}$ fragmentation process, and $\overline{K}^{0,-}$ represents a single $\overline{K}^0$ or $K^-$ from $c\overline{c}$ fragmentation required to assure overall balance of strangeness in the event. The photon from the decay $D_s^{*+} \to D_s^+\gamma$ is referred to as the signal photon.

Event selection begins with $\overline{D}_{\rm TAG}$ construction. Candidates are reconstructed in the following modes: $\overline{D}^0 \to K^+\pi^-(\pi^0)$, $K^+\pi^-\pi^-\pi^+(\pi^0)$, or $K_S^0\pi^+\pi^-(\pi^0)$, and $D^- \to K^+\pi^-\pi^-(\pi^0)$, $K_S^0\pi^-(\pi^0)$, or $K_S^0\pi^-\pi^-\pi^+$. The $\chi^2$ probability for the geometric vertex fit of the TAG decay products must exceed 0.1%. The minimum required CM momentum of the $\overline{D}_{\rm TAG}$ candidate is 2.35 GeV/$c$. It is chosen near the kinematic limit for charm mesons arising from $B$ decays in order to eliminate the associated large combinatoric background. The mass of the $\overline{D}_{\rm TAG}$ candidate must lie in the range 1.7-2.1 GeV/$c^2$.

A single $K^-$ or $K_S^0$ from $c\overline{c}$ fragmentation that does not have tracks in common with the $\overline{D}_{\rm TAG}$ combination is found. Kaons are identified using information from the DCH and DIRC. A $K_S^0$ candidate is reconstructed through its decay to two charged pions which must originate from a common vertex. The dipion invariant mass must be within 25 MeV/$c^2$ of the nominal $K_S^0$ mass value [9], and the flight distance must be greater than three times its resolution. Neutral pions are reconstructed through their decay to two photons; the invariant mass of the photon pair must be within 10 MeV/$c^2$ of the nominal $\pi^0$ mass value [9]. Any charged or neutral pion not associated with the $\overline{D}_{\rm TAG}$ or the fragmentation kaon is assigned to the fragmentation $X$ candidate.

A $D_s^{*+}$ candidate is reconstructed as the missing particle with its four-momentum defined as, $P_{D_s^{*+}} = P_{e^+e^-} - (P_{\overline{D}_{\rm TAG}} + P_{K_S^0/K^-} + P_X)$, where the four-momenta ($P$) are from the initial state, $\overline{D}_{\rm TAG}$, the fragmentation kaon and the fragmentation $X$, respectively. The mass of the $D_s^{*+}$ candidate must be within 200 MeV/$c^2$ of the nominal $D_s^{*+}$ mass value [9]. The production vertex of surviving candidates is then fitted using mass, energy and collision point constraints.

In order to be consistent with the decay sequence $D_s^+ \to \tau^+\nu_\tau$, $\tau^+ \to e^+\nu_e\overline{\nu}_\tau$, a $D_s^+$ candidate is selected by requiring that there be a single $e^+$ in the event. The $e^+$ must have a minimum number of coordinates in the SVT and the DCH to ensure a good quality track. Similarly, the decay $D_s^+ \to K_S^0K^+$ is identified by requiring a single $K_S^0$ and $K^+$ pair. In addition, these candidates must not have tracks in common with $\overline{D}_{\rm TAG}$ or the fragmentation kaon.

The four-momentum of the $D_s^+$ candidate, for both the signal and the normalization mode, is defined as the recoil in the two-body decay $D_s^{*+} \to D_s^+\gamma$, $P_{D_s^+} = P_{D_s^{*+}} - P_\gamma$, where $P_\gamma$ is the four-momentum of the signal photon candidate, which must have CM energy greater than 100 MeV. The resulting $D_s^+$ candidate must have mass within 200 MeV/$c^2$ of the nominal $D_s^+$ mass value [9].

Surviving $D_s^+$ candidates are further separated from background by requiring that the $\chi^2$ probability for the $D_s^{*+}$ kinematic vertex fit exceed 0.1%, and that the CM momentum of the $D_s^+$ candidate exceed 3.0 GeV/$c$. The whole reconstruction procedure was evaluated using `GEANT4`-based [18] Monte Carlo (MC) events generated with `EvtGen` [19]. The generated MC samples for $D_s^+ \to \tau^+\nu_\tau$ ($\tau^+ \to e^+\nu_e\overline{\nu}_\tau$), $D_s^+ \to K_S^0K^+$ and $c\overline{c}$ correspond to 14, 26 and 2 times the acquired data samples, respectively.

After the final selection, the only background decay modes which contribute to the peak at the recoil $D_s^+$ mass (with the expected yields and shape determined from MC events weighted to 427 fb$^{-1}$) are $D_s^+ \to \eta e^+\nu_e$ (226 events), $D_s^+ \to \eta' e^+\nu_e$ (24 events), $D_s^+ \to \phi e^+\nu_e$ (75 events) and $D_s^+ \to K_L^0 e^+\nu_e$ (59 events).

The yields for the signal and normalization mode are determined from unbinned maximum-likelihood fits to the respective recoil $D_s^+$ mass and extra energy ($E_{\rm extra}$) distributions. As described earlier, the recoil $D_s^+$ four-momentum is defined as $P_{D_s^+} = P_{D_s^{*+}} - P_\gamma$; $E_{\rm extra}$ is reconstructed as the sum of the CM energy of all photons in the event with laboratory energy greater than 30 MeV which are not associated with any of the reconstructed charged-particle tracks or reconstructed neutral pions of the event. The signal photon is also excluded. The value of $E_{\rm extra}$ has been found to discriminate most effectively between signal and background events when it is required to be in the range 0-0.5 GeV. The distributions of $E_{\rm extra}$ for signal MC, background MC (events passing selection that do not include the signal) and data are shown in Fig. 1. The difference between data and MC at $E_{\rm extra} = 0$ is due to MC under-estimation of beam-related backgrounds and of noise in the calorimeter. It has been verified that the MC gives a a good description of the $E_{\rm extra}$ distribution in data for values of $E_{\rm extra}$ above 20 MeV. Correlations between the recoil $D_s^+$ mass and $E_{\rm extra}$ were found to be negligible. Due to the discontinuity in the $E_{\rm extra}$ distribution, the data are divided into two samples from which the results are determined using a simultaneous unbinned maximum likelihood fit. For $E_{\rm extra} = 0$, only the recoil $D_s^+$ mass is used as a discriminating variable. For $E_{\rm extra} > 0$, $E_{\rm extra}$ and the recoil $D_s^+$ mass are used. The fit components are signal, peaking background and non-peaking background. The $D_s^+ \to K_S^0K^+$ mode is found to have no peaking background in MC, and thus no peaking background component is included in the fit.

The signal recoil $D_s^+$ mass probability density function (PDF) consists of a bifurcated Gaussian function with a tail component (BFG) [20], plus a Novosibirsk function [21]. The shape of the $E_{\rm extra}$ distribution for background is taken from the data sidebands of the recoil $D_s^+$ mass distribution ($M_{\rm Recoil} < 1.95$ GeV/$c^2$ and $M_{\rm Recoil} > 2.0$ GeV/$c^2$). The remaining PDFs are empirical functions which describe the MC predictions and data. The background recoil mass PDF is a Novosibirsk function. The peaking background recoil mass PDF is a BFG plus a

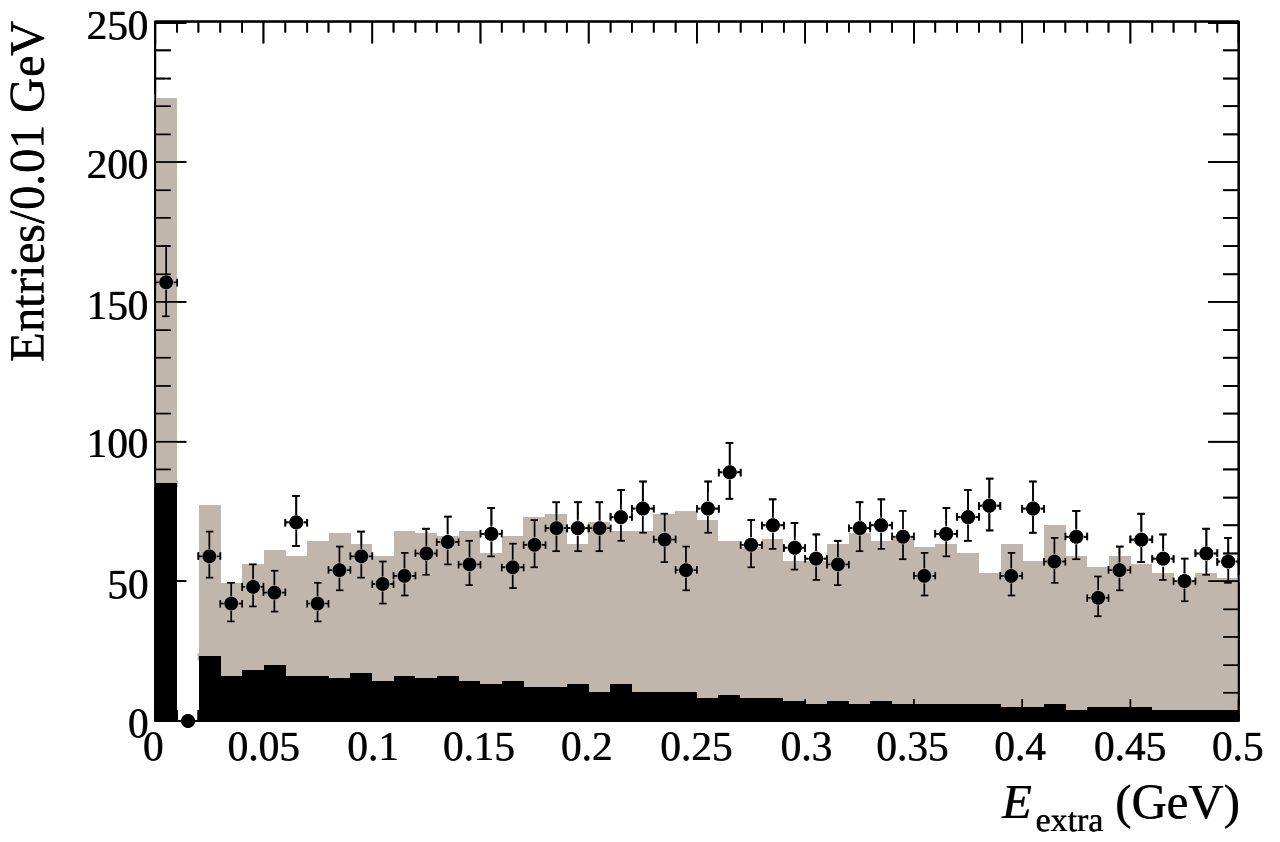

FIG. 1: Extra energy ($E_{\rm extra}$) in MC and data for $D_s^+ \to \tau^+\nu_\tau$ ($\tau^+ \to e^+\nu_e\overline{\nu}_\tau$). The MC is normalized to the size of the data sample. The points represent data, the solid grey histogram is from background MC and the solid black histogram is from signal MC. The gap between 0 and 20 MeV is due to the minimum energy requirement on photon candidates.

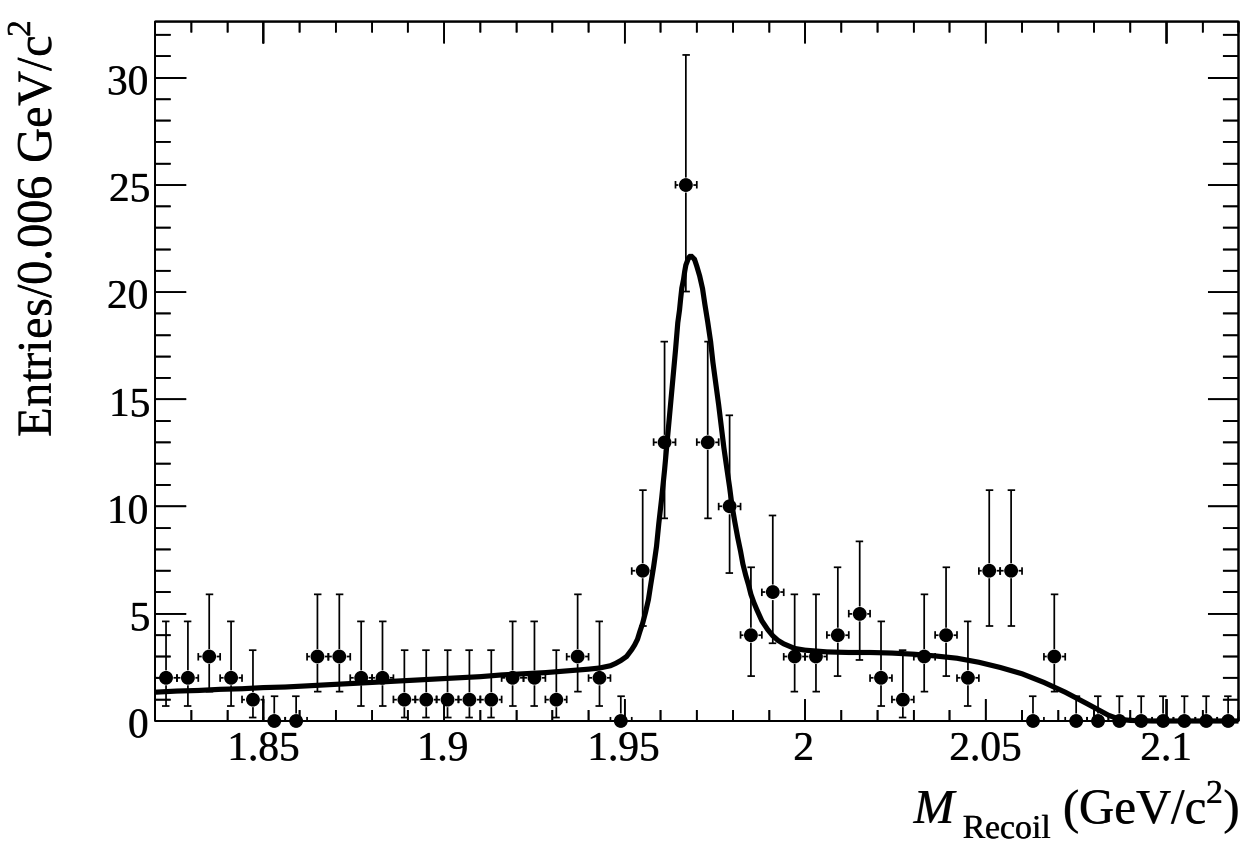

FIG. 2: $D_s^+$ candidate recoil mass for $D_s^+ \to \tau^+\nu_\tau$ ($\tau^+ \to e^+\nu_e\overline{\nu}_\tau$) with $E_{\rm extra} = 0$; the curve results from the fit described in the text.

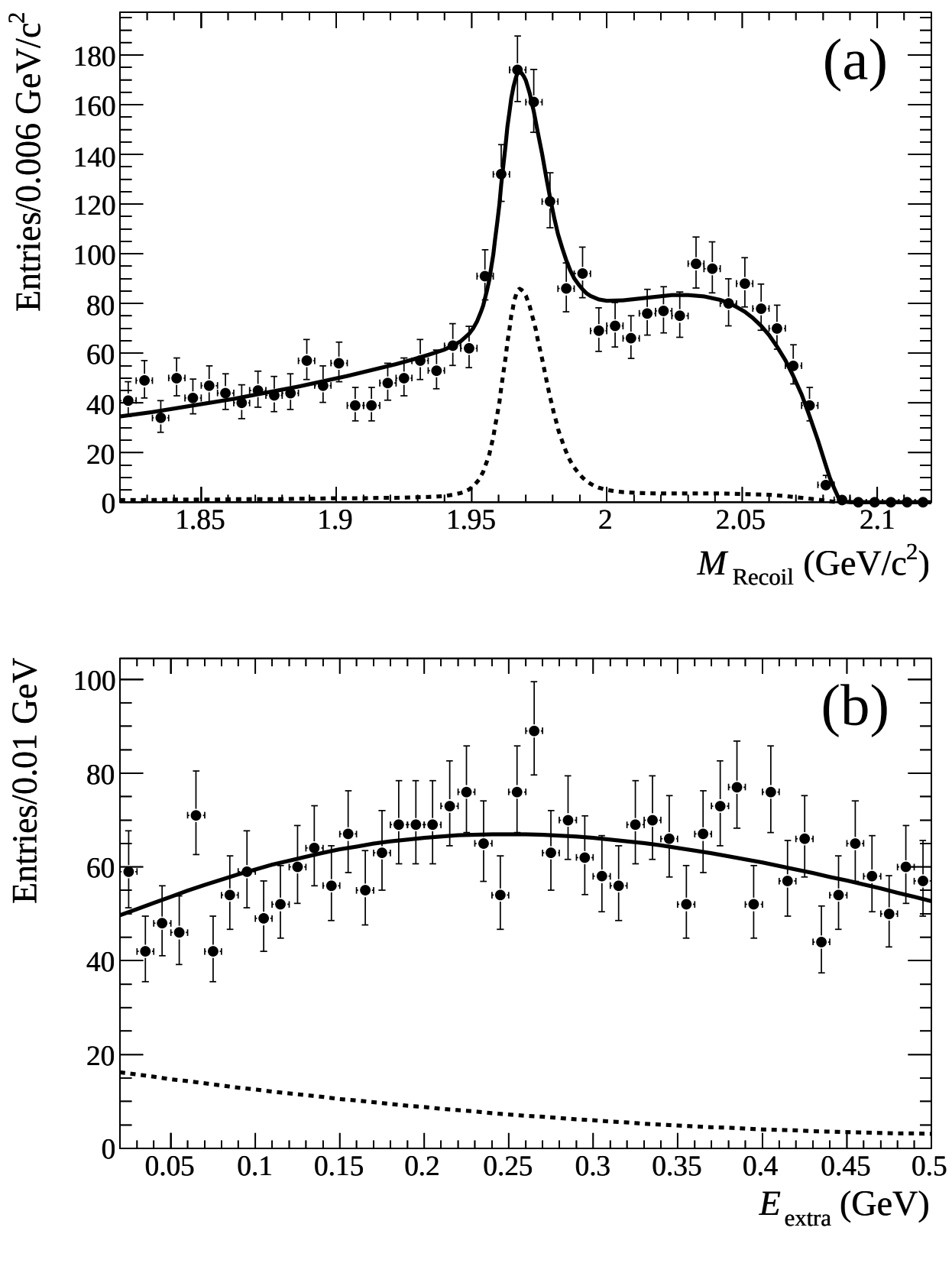

FIG. 3: (a) $D_s^+$ candidate recoil mass for $D_s^+ \to \tau^+\nu_\tau$ ($\tau^+ \to e^+\nu_e\overline{\nu}_\tau$) with $E_{\rm extra} > 0$. (b) $E_{\rm extra}$ for $E_{\rm extra} > 0$. The solid curves result from the fit described in the text, and the dashed curves represent the signal contribution.

Novosibirsk function. Second order polynomial functions are used for the signal and peaking background PDFs for $E_{\rm extra}$. A Novosibirsk function is used for the background $E_{\rm extra}$ PDF. The parameters describing the shape of each PDF are obtained from fits to MC distributions. The systematic uncertainties introduced by this procedure are discussed below. Only the parameters specifying the number of signal and background events are allowed to vary in the fit. The number of peaking background events is determined from the MC normalized to the data sample. Using MC pseudoexperiments, the fitting procedure is found to yield unbiased estimates of the signal yield. The fits to data are shown in Figs. 2–5, and yield $N_S^{\tau\nu_\tau}$=448±36 events and $N_S^{K_S^0K^+}$=333±28 events. Using Eq. (2) and the total efficiencies ($\epsilon^{\tau\nu} = 0.075\%$, $\epsilon^{K_S^0K^+} = 0.044\%$) the branching fraction for $D_s^+ \to \tau^+\nu_\tau$ is measured to be (4.5±0.5)%, where the uncertainty is statistical only.

The systematic uncertainties associated with the selection criteria are evaluated by comparing the selection efficiencies for MC and data separately for each selection criterion. The selection efficiency is defined as $N_{N-1}/N_{\rm All}$ where $N_{N-1}$ is the number of events passing all selection criteria except the one being evaluated, and $N_{\rm All}$ is the number of events passing all selection criteria. The uncertainty is then defined as $|1 - R_{D_s^+\to\tau^+\nu_\tau}/R_{D_s^+\to K_S^0K}|$ where $R$ is the ratio of data and MC efficiencies for each decay mode. The uncertainty associated with the $\chi^2$ probability for the kinematic vertex fit is 1.1%, and that with the CM momentum of the $D_s^+$ meson is 2.7%. The uncertainties in the PDF distributions are evaluated by individually varying each PDF parameter by one standard deviation and obtaining the corresponding variation in the fitted number of signal and background events. To obtain the total uncertainty the indi-vidual contributions are added in quadrature. The impact of the uncertainty in the number of peaking background events for $D_s^+ \to \tau^+\nu_\tau$ on the signal yield is assessed by varying the uncertainties for the individual branching fractions, and refitting for the number of signal events. The peaking background modes and their branching fractions are: $D_s^+ \to \eta e^+\nu_e$ (2.9±0.6)%, $D_s^+ \to \eta' e^+\nu_e$ (1.02±0.33)%, $D_s^+ \to \phi e^+\nu_e$ (2.36±0.26)% and $D_s^+ \to K_L^0 e^+\nu_e$ (0.19±0.05)% [9]. Other sources of systematic uncertainty include tracking efficiency (0.34% per track) and $e^+$ identification efficiency (0.82%). The uncertainties from tagging and fragmentation particles cancel in the ratio of the signal and reference modes. Table I summarizes the systematic uncertainty estimates on the branching fraction.

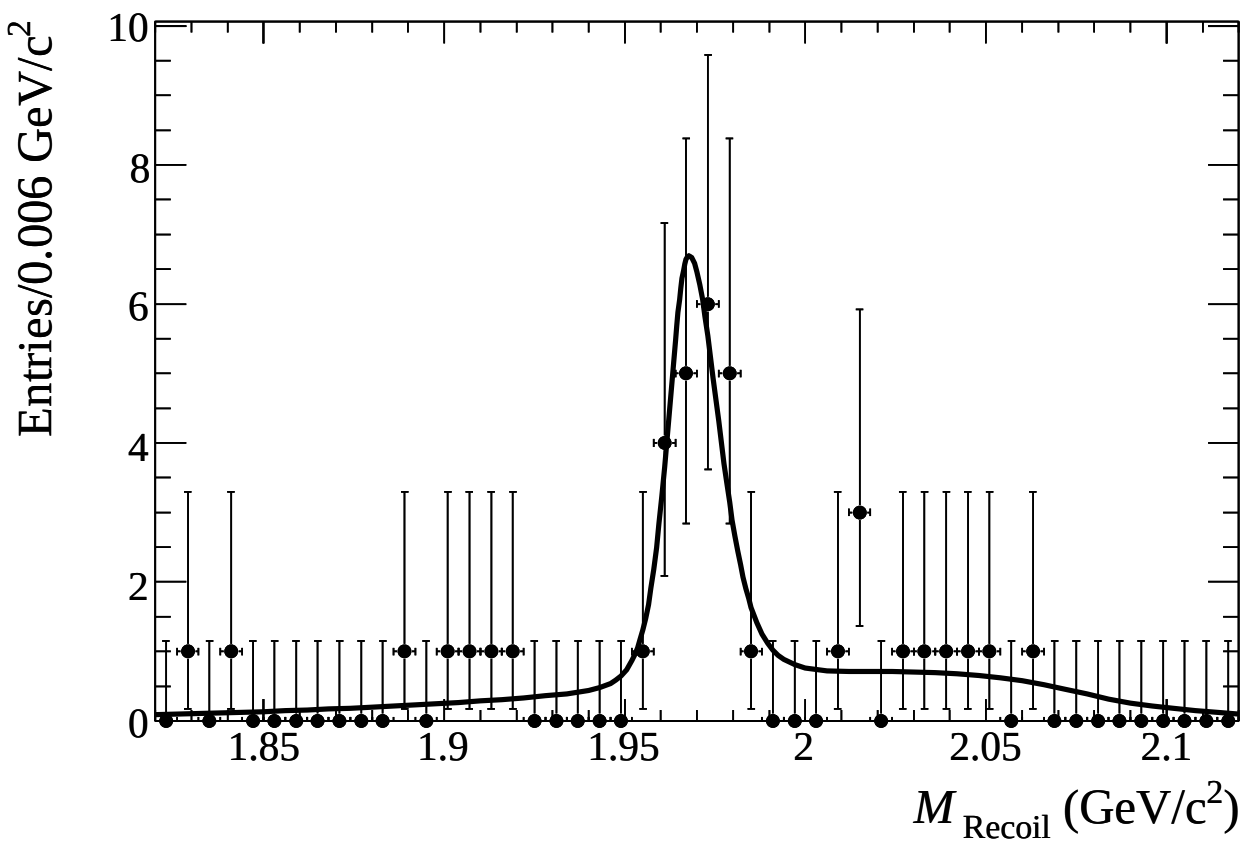

FIG. 4: $D_s^+$ candidate recoil mass for $D_s^+ \to K_S^0 K^+$ ($K_S^0 \to \pi^+\pi^-$) with $E_{\rm extra} = 0$; the curve results from the fit described in the text.

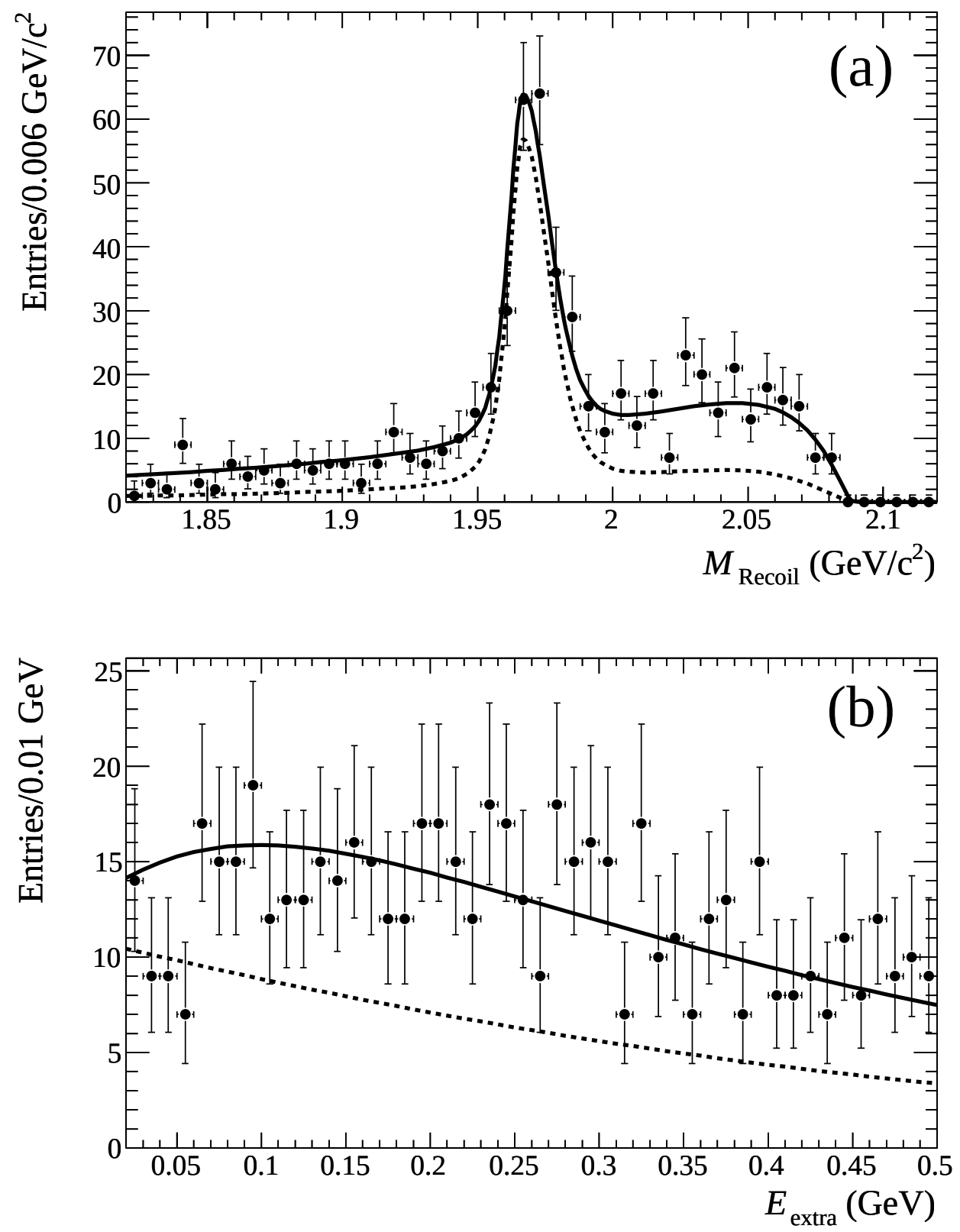

FIG. 5: (a) $D_s^+$ candidate recoil mass for $D_s^+ \to K_S^0 K^+$ ($K_S^0 \to \pi^+\pi^-$) with $E_{\rm extra} > 0$. (b) $E_{\rm extra}$ for $E_{\rm extra} > 0$. The solid curves result from the fit described in the text, and the dashed curves represent the signal contribution.

TABLE I: Relative systematic uncertainty estimates on the branching fraction.

| Source | Uncertainty (%) |
|---|---|
| Event Selection | 3.0 |
| Particle Identification | 0.82 |
| Tracking | 0.68 |
| $\tau\nu_\tau$ PDF Distribution | +7.7 -4.7 |
| $K_S^0 K$ PDF Distribution | +4.9 -0.6 |
| Peaking Background | +4.5 -4.3 |

In conclusion, using an integrated luminosity of 427 fb$^{-1}$ collected with the *BABAR* detector, the branching fraction for the decay $D_s^+ \to \tau^+\nu_\tau$ is measured to be (4.5±0.5±0.4±0.3)%, where the first uncertainty is statistical, the second systematic and the third from the uncertainties on the branching fractions for $D_s^+ \to K_S^0 K^+$, $K_S^0 \to \pi^+\pi^-$ and $\tau^+ \to e^+\nu_e\overline{\nu}_\tau$ [9]. The decay constant is extracted using Eq. (1) and the values of $m_\tau$(1776.84±0.17 MeV/$c^2$), $m_{D_s}$(1968.49±0.34 MeV/$c^2$), $\tau_{D_s}$(0.500±0.007 ps) and by assuming $|V_{cs}|$ = $|V_{ud}|$ = 0.97425±0.00022 [22]. The value obtained is $f_{D_s}$ = (233±13±10±7) MeV, where the first uncertainty is statistical, the second is systematic, and the third is from the uncertainties on the external measured quantities used in the calculation [9]. The results presented here agree to within one standard deviation with the most recent CLEO-c result for $D_s^+ \to \tau^+\nu_\tau$ [10, 11] and recent unquenched lattice QCD calculations of $f_{D_s}$ [5, 7, 8].

We are grateful for the excellent luminosity and machine conditions provided by our PEP-II colleagues, and for the substantial dedicated effort from the computing organizations that support *BABAR*. The collaborating institutions wish to thank SLAC for its support and kind hospitality. This work is supported by DOE and NSF (USA), NSERC (Canada), CEA and CNRS-IN2P3 (France), BMBF and DFG (Germany), INFN (Italy), FOM (The Netherlands), NFR (Norway), MES (Russia), MEC (Spain), and STFC (United Kingdom). Individuals have received support from the Marie Curie EIF (European Union) and the A. P. Sloan Foundation.

* Now at Temple University, Philadelphia, Pennsylvania 19122, USA
† Also with Università di Perugia, Dipartimento di Fisica, Perugia, Italy
‡ Also with Università di Roma La Sapienza, I-00185 Roma, Italy
§ Now at University of South Alabama, Mobile, Alabama 36688, USA
¶ Also with Laboratoire de Physique Nucléaire et de Hautes Energies, IN2P3/CNRS, Université Pierre et Marie Curie-Paris6, Université Denis Diderot-Paris7, F-75252 Paris, France
** Also with Università di Sassari, Sassari, Italy
†† Deceased